\documentclass[amsfonts,amsmath,tightenlines,preprintnumbers,nofootinbib,superscriptaddress]{revtex4}
\usepackage{graphicx}
\usepackage{epsfig}
\usepackage{bm}

\usepackage{epsfig}

\newcommand{\Choose}[2]{{\begin{pmatrix} {#1} \\ {#2} \end{pmatrix}}}

\def\pislash{ {\pi\hskip-0.6em /} }

\def\nopi{ {\rm EFT}(\pislash) }

\begin{document}

\preprint{NT@UW-08-01}

\begin{figure}[!t]
\vskip -1.1cm
\leftline{
{\epsfxsize=1.2in \epsfbox{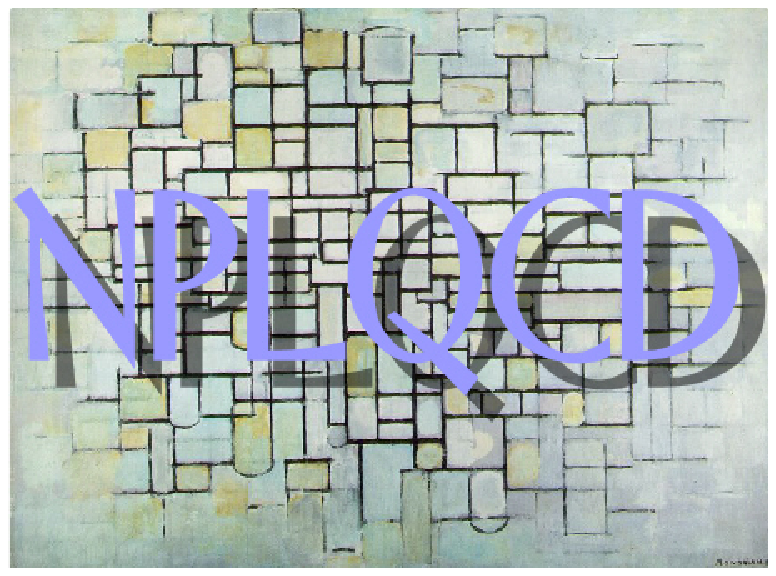}}}
\vskip 1.5cm
\end{figure}

\title{The Energy of $n$ Identical Bosons in a Finite Volume at ${\cal O}(L^{-7})$}

\author{William Detmold and Martin J. Savage} \affiliation{Department of Physics,
  University of Washington, Box 351560, Seattle, WA 98195, USA}

\date\today

\begin{abstract}
  The volume dependence of the ground-state energy of $n$ identical
  bosons with short-range interactions in a periodic spatial volume
  with sides of length $L$ is calculated at order $L^{-7}$ in the
  large volume expansion.  This result will enable a refined
  determination of the $\pi^+\pi^+\pi^+$ interaction from lattice QCD
  calculations.
\end{abstract}

\maketitle

It is now well established that two-body interactions between hadrons can be studied
with lattice  QCD as
the volume dependence of the energy spectrum of two hadrons is related to their
scattering amplitude below inelastic thresholds~\cite{Luscher:1986pf,Luscher:1990ux}.
Recently, this method has been used to 
determine the $\pi^+\pi^+$ scattering length~\cite{Beane:2007xs},
$a_{\pi^+\pi^+}$, with $\sim 1\%$ precision with a  $n_f=2+1$  fully-dynamical 
mixed-action lattice QCD calculation.
In order to extract the many-body interactions from lattice QCD calculations, the
energy of multi-hadron states in a finite volume must be calculated with
lattice QCD and 
combined with the known dependence of this energy on the many-body interactions.
The ground state energy of a system of $n$ identical
bosons with short-range interactions in a cubic volume with sides of length $L$
was recently computed at ${\cal O}(L^{-6})$ in the large-volume expansion~\cite{Beane:2007qr}.
The underlying motivation for that work, 
which builds upon the classic works of 
Refs.~\cite{Huang:1957im,Wu:1959,HandP,Sawada}, was to provide a way to determine the three-body interactions between
$\pi^+$'s from lattice QCD calculations, which first enters at that order~\cite{Beane:2007qr}.
In Ref.~\cite{Beane:2007es},
this result was used in conjunction with  lattice QCD calculations of 
multi-pion systems to
determine the interaction between three $\pi^+$'s for the
first time.
In order to refine the determination  of the $\pi^+\pi^+\pi^+ $ interaction, 
here we compute the contribution to the energy-shift of $n$ identical bosons 
at ${\cal O}(L^{-7})$ in the large volume expansion.
The energy-shift of three identical bosons in a  finite-volume has been been
computed recently in Ref.~\cite{Tan:2007bg}, and our $n=3$ calculation agrees.

The ground-state energy of $n$ identical bosons is calculated 
using standard Schr\"odinger perturbation theory, 
with a Hamiltonian, appropriate to the order we are working in the large volume
expansion, of the form
\begin{eqnarray}
H & = & 
\sum_{\bf k}\ h_{\bf k}^\dagger \  h_{\bf k}\ \left(\ {|{\bf k}|^2\over 2 M}\ -\
  {|{\bf k}|^4\over 8 M^3}\ \right)
\nonumber\\
&& + \
{1\over (2!)^2}
\sum_{\bf Q,k,p}\  h_{\bf {Q\over 2}+k}^\dagger  h_{\bf {Q\over 2}-k}^\dagger\
h_{\bf {Q\over 2}+p}\ h_{\bf {Q\over 2}-p}
\ \left(\ {4\pi a\over M}\ +\ {\pi a\over M}\left( a r - {1\over 2 M^2}\right)
  \left(\  |{\bf k}|^2 + |{\bf p}|^2\ \right)\ \right)
\nonumber\\
&& + \
{\eta_3(\mu)\over (3!)^2}\ 
\sum_{\bf Q,k,p,r,s}\  h_{\bf {Q\over 3}+k}^\dagger  h_{\bf {Q\over 3}+p}^\dagger\ h_{\bf {Q\over 3}-k-p}^\dagger\
 h_{\bf {Q\over 3}+r}  h_{\bf {Q\over 3}+s}\ h_{\bf {Q\over 3}-r-s}
\ ,
\label{eq:interaction}
\end{eqnarray}
where the operator $h_{\bf k}$ annihilates a $\pi^+$ with momentum ${\bf k}$
with unit amplitude. The divergences that arise at loop-level are regulated
with dimensional regularization, and therefore the coefficients of the two-body
interaction can be readily identified with the parameters describing
the scattering amplitude: the scattering length, $a$, and the effective range,
$r$ \ ($p\cot\delta=-{1\over a}\ +\ {1\over 2} r
  p^2 + \ ... $\ ).
The terms proportional to $M^{-3}$ in eq.~(\ref{eq:interaction})
describe  the leading effects of
relativity.
Only the momentum independent
three-body interaction,  $\eta_3(\mu)$, is required at ${\cal O}(L^{-7})$.
Our method of computation 
is equivalent to the 
pionless EFT describing low-energy  nucleon-nucleon interactions, 
$\nopi$~\cite{Kaplan:1998tg,van Kolck:1998bw,Chen:1999tn} (when modified to
describe systems with natural scattering lengths) 
and the method of pseudo-potentials used in our previous work.  
The divergences that
occur in loop diagrams are renormalized order-by-order in the
expansion, preserving the power counting,
and hence the explicit dependence of the bare three-body
coefficient on the renormalization scale, $\mu$.

The calculation of the energy-shift of $n$ identical bosons at 
${\cal O}(L^{-7})$ due to the interactions defined in eq.~(\ref{eq:interaction}) 
is straightforward but tedious.
We will not delve into the details, referring the reader to our previous
work~\cite{Beane:2007qr} and that of Ref.~\cite{Tan:2007bg}, and simply state the result.
The energy-shift of the ground state is
\begin{eqnarray}
\label{eq:7}
  E_0(n,L) &=&
  \frac{4\pi\, a}{M\,L^3}\Choose{n}{2}\Bigg\{1
-\left(\frac{a}{\pi\,L}\right){\cal I}
+\left(\frac{a}{\pi\,L}\right)^2\left[{\cal I}^2+(2n-5){\cal J}\right]
\nonumber 
\\&&\hspace*{2cm}
-
\left(\frac{a}{\pi\,L}\right)^3\Big[{\cal I}^3 + (2 n-7)
  {\cal I}{\cal J} + \left(5 n^2-41 n+63\right){\cal K}\Big]
\nonumber
\\&&\hspace*{2cm}
+
\left(\frac{a}{\pi\,L}\right)^4\Big[
{\cal I}^4 - 6 {\cal I}^2 {\cal J} + (4 + n - n^2){\cal J}^2 
+ 4 (27-15 n + n^2) {\cal I} \ {\cal K}
\nonumber\\
&&\hspace*{4cm}
+(14 n^3-227 n^2+919 n-1043) {\cal L}\ 
\Big]
\Bigg\}
\nonumber
\\&&\hspace*{0cm}
+\Choose{n}{2} \frac{8\pi^2 a^3 r }{M\, L^6}\ 
\Big[\  1\ +\ \left(\frac{a}{\pi\,L}\right) 3(n-3) {\cal I}\ 
\Big]
\nonumber\\
&&
+\Choose{n}{3} {1\over L^6}\ 
\left[\ 
\eta_3(\mu)\ +\ {64\pi a^4\over M}\left(3\sqrt{3}-4\pi\right)\ \log\left(\mu
  L\right)\ -\ 
{96 a^4\over\pi^2 M} \ {\cal S}
\ \right]
\left[ 1\ -\ 6 \ \left({a\over \pi L}\right) \ {\cal I} \ \right]
\nonumber\\
&&
+\Choose{n}{3}\left[\ 
{192 \ a^5\over M\pi^3 L^7} \left( {\cal T}_0\ +\ {\cal T}_1\ n \right)
\ +\ 
{6\pi a^3\over M^3 L^7}\ (n+3)\ {\cal I}\ 
\right]
\ \ + \ {\cal O}\left(L^{-8}\right)
\ \ \ \ .
\label{eq:energyshift}
\end{eqnarray}
where the geometric constants that enter are~\footnote{
The constants ${\cal I, J, K}$ were defined previously in
Ref.~\cite{Beane:2007qr}, while the constant ${\cal L}$ is defined to be 
the integer-triplet sum
\begin{eqnarray}
{\cal L}& = & \sum_{{\bf n}\ne{\bf 0}}{1\over |{\bf n}|^8}
\ \ ,
\nonumber
\end{eqnarray}
and is equal to ${\cal L}=\alpha_4$ in the notation of Ref.~\cite{Tan:2007bg}.
The constants ${\cal T}_{0,1}$ arise from combinations of 
up to three-loop diagrams, and involve three-, six- and nine-dimensional sums over
integers, and can be written in terms of 
constants defined in Ref~\cite{Tan:2007bg} plus one additional sum, $S_1$,
\begin{eqnarray}
{\cal T}_0 + {\cal T}_1\ n & = & 
{1\over 4}\alpha_{1AA1} - {\cal I} \ \alpha_{1A1}
+{1\over 2}(2n-9)\alpha_{2A1} 
+ {3\over 4}(n-4) \alpha_{1B1}
- {1\over 4} (7n-29) {\cal L}
+2(n-3) S_1
\ \ \ ,
\nonumber
\end{eqnarray}
where
\begin{eqnarray}
S_1 & = & \sum_{{\bf n},{\bf j} \ne {\bf 0}}
{1\over |{\bf n}|^2 |{\bf j}|^4 
\left[\ |{\bf n}|^2 + | {\bf n} + {\bf j}|^2\right]}
\ =\ 92.42215
\ \ \ .
\nonumber
\end{eqnarray}
}
\begin{align} 
&{\cal I}\ =\  -8.9136329
&
{\cal T}_0\  = -4116.2338
\nonumber\\
&{\cal J}\ =\  16.532316
&
{\cal T}_1\  = \ 450.6392
\nonumber\\
&{\cal K}\  = \  8.4019240
&
{\cal S}_{\rm MS}\ = \ -185.12506
\nonumber\\
&{\cal L}\  = \  6.9458079
&
\label{eq:sums}
\end{align}
and ${\tiny \Choose{n}{k}}$=$n!/(n-k)!/k!$.  
The last term in the last bracket of eq.~(\ref{eq:energyshift}) is the
leading relativistic contribution to the energy-shift. 
Deviations from the energy-shift of $n$-bosons computed
with non-relativistic quantum mechanics arise only for  three or more
particles as the two-particle energy-shift has the same form when computed in 
non-relativistic quantum mechanics and in quantum field
theory~\cite{Luscher:1986pf,Luscher:1990ux}. 
In eq.~(\ref{eq:sums}), ${\cal S}_{\rm MS}$ is the value of
the scheme-dependent quantity
${\cal S}$ in the Minimal Subtraction (MS) scheme that we have 
employed to renormalize the theory (a change in scheme results in a change in 
${\cal S}$ and a compensating change in $\eta_3(\mu)$).\footnote{
In the notation of Ref.~\cite{Beane:2007qr},
${\cal S}_{\rm MS} = 2{\cal  Q}+{\cal R}$.
The numerical value in eq.~(\ref{eq:sums}) corrects a minor error in ${\cal Q}$
in Ref.~\cite{Beane:2007qr}.}
The ${\cal T}_i$ are renormalization scheme independent.
Our result at $n=2$ agrees with large volume expansion of Ref.~\cite{Luscher:1986pf,Luscher:1990ux},
and at $n=3$ agrees with the previous computation by Shina Tan~\cite{Tan:2007bg}.

The renormalization-scale independent, but volume dependent, quantity
\begin{align} 
\overline{\eta}_3^L & =  
\eta_3(\mu)\ +\ {64\pi a^4\over m}\left(3\sqrt{3}-4\pi\right)\ \log\left(\mu
  L\right)\ -\ 
{96 a^4\over\pi^2 m} {\cal S}
\label{eq:etathreebar}
\end{align}
was determined in recent lattice QCD calculations~\cite{Beane:2007es}.
It was found to be non-vanishing in systems of three, four and five $\pi^+$'s
at pion masses of $m_\pi\sim 290$ and $350~{\rm MeV}$ in  a $\sim (2.5~{\rm
  fm})^3$ volume, when extracted at ${\cal O}(L^{-6})$ in the large volume expansion.
Its size was found to be consistent with expectations based upon naive
dimensional analysis, $\overline{\eta}_3^L\sim 1/(m_\pi f_\pi^4)$.
Our result will allow for further refinement of such extractions.

\acknowledgements{We thank Silas Beane and Shina Tan for useful discussions. 
This work is supported by the Department of Energy under grant DE-FG03/974014. }

\end{document}